\documentclass[amsmath,amssymb,article,onecolumn]{revtex4}
\usepackage[english]{babel}
\usepackage{natbib}
\usepackage{multirow}

\usepackage{rotating} 

\usepackage{color}

\usepackage{framed}

\begin{document}

\title{The new Resonating Valence Bond Method for ab-initio Electronic Simulations}

\author{Sandro Sorella} 
\affiliation{
Scuola Internazionale Superiore di Studi Avanzati (SISSA) and Democritos National Simulation Center, Istituto Officina dei Materiali del CNR, via Bonomea 265, 34136 Trieste, Italy}
\email{sorella@sissa.it}

\author{Andrea Zen} 
\affiliation{Dipartimento di Fisica, Universit\`a di Roma ``La Sapienza'', Piazzale Aldo Moro 2, 00185 Rome, Italy} 
\email{andrea.zen@uniroma1.it}

\begin{abstract}
The Resonating Valence Bond theory of the chemical bond was introduced soon after the discovery of quantum mechanics and has contributed to explain the role of electron correlation within a particularly
simple and intuitive approach where the chemical bond between two nearby atoms  
is described by one or more singlet electron pairs.
In this chapter Pauling's resonating valence bond theory of the chemical bond 
is revisited within a new formulation,   introduced 
by P.W. Anderson after the discovery  of High-Tc superconductivity.  
It is shown that this intuitive picture 
of electron correlation becomes now practical and efficient, 
since it allows us to  faithfully exploit the locality of the 
electron correlation, and to describe several new phases of matter, such as Mott insulators,
High-Tc superconductors, and spin liquid phases.
\end{abstract}

\maketitle

\section{Introduction}
\label{sec:1}
Soon after  the discovery of quantum mechanics Linus Pauling introduced the so called 
  "Resonating Valence Bond" (RVB)  theory of the chemical bond, an innovative 
point of view  that 
soon became popular  and certainly had  a  tremendous impact in chemical  and 
physical sciences. Until now this theory remains extremely useful  to explain the role of electron correlation within a particularly
simple and intuitive approach, where the chemical bond is described by a main ingredient: the  singlet formed by electron pairs belonging to different atoms, when they  become sufficiently close.
It was soon realized however that this beautiful description was not practical simply because  the number of resonating chemical bonds increases exponentially with the number of atoms. Thus, the molecular orbital
approach, the basic ingredient of Hartree-Fock, DFT and post Hartree-Fock methods,
became the standard recipe for the electronic simulation
as it is discussed in several chapters of this book (
(see in particular Chap. ``Tensor Product Approximation (DMRG) and Coupled Cluster Method in Quantum Chemistry'' by Legeza et al., ``Computational Techniques for Density Functional-based Molecular Dynamics Calculations in Plane-Wave and Localized Basis Sets'' by Tzanov and Tuckerman, and Chap. ``Application of (Kohn-Sham) Density Functional Theory to Real Materials'' by Ghiringhelli).
We  show in this chapter that a new resonating valence bond  scheme is possible.  \footnote{Google TurboRVB web page for further informations} 
  This new formulation of the resonating valence bond theory
is borrowed by the fascinating theory of High-Tc superconductivity
proposed by Anderson, right after its discovery in the 90's.
By means of quantum Monte Carlo computations and the increasing power of modern supercomputers this intuitive picture
of electron correlation becomes  practical and efficient,
since it allows us to exploit the short range nature of the
chemical bond, as well as, to describe several new phases of matter, such as Mott insulators
and High-Tc superconductors.

When two atoms are close their valence electrons interact with each other, forming the chemical bond.
In the most simplified picture of two single valence electrons (e.g., in the $H_2$ molecule) 
the singlet pairing allows us to minimize the Coulomb energy and satisfy the antisymmetry of the 
wave function by acting only on its spin part 
$ { 1 \over \sqrt{2} } ( | \uparrow \rangle | \downarrow  \rangle - | \downarrow \rangle | \uparrow \rangle ) $, 
whereas its orbital part $f$ is symmetric and 
is a generic function of the two valence electron coordinates:
\begin{equation}
f(\vec r, \vec r^\prime) = f (\vec r^\prime,\vec r) ,
\end{equation}  
being non-zero even in the interatomic region where the chemical bond is formed.  
As a matter of fact, a  wave function of the above form  is {\em exact }
for two electrons in a singlet ground state and allows us to describe, for instance, the 
 $H_2$ molecule at all interatomic distances, a well-known case 
where a restricted Hartree-Fock wave function 
\begin{equation}
f_{HF}(\vec r, \vec r^\prime) = \phi (\vec r) \phi (\vec r^\prime) 
\end{equation} 
miserably fails at large distances, because it is not able to reproduce the Heitler-London 
solution (HLS):
\begin{equation} \label{hl}
f_{HL}(\vec r, \vec r^\prime) = \phi_A (\vec r) \phi_B (\vec r^\prime) + \phi_B (\vec r) \phi_A (\vec r^\prime) ,
\end{equation} 
where $\phi$ indicates an   Hartree-Fock  molecular orbital, with $\phi_A$ ($\phi_B$) 
an atomic one  localized around the atom $A$ ($B$). 

The simple extension of this simple singlet valence bond of two electrons was originally 
formulated in terms of simple Slater determinants, e.g. two determinants for the 
 single bond in $H_2$. Indeed the simple superposition of the bonding $\phi_+ = \phi_A+\phi_B$ 
and the antibonding $\phi_- = \phi_A-\phi_B$ orbitals in the  pairing function
\begin{equation}
f(\vec r, \vec r^\prime) =  \phi_+ (\vec r) \phi_+(\vec r^\prime) + \lambda \phi_- (\vec r) \phi_- (\vec r^\prime) 
\end{equation} 
allows us  to recover the HLS for $\lambda=-1$ at large distance.

\begin{figure}[h!]
\includegraphics[scale=.45]{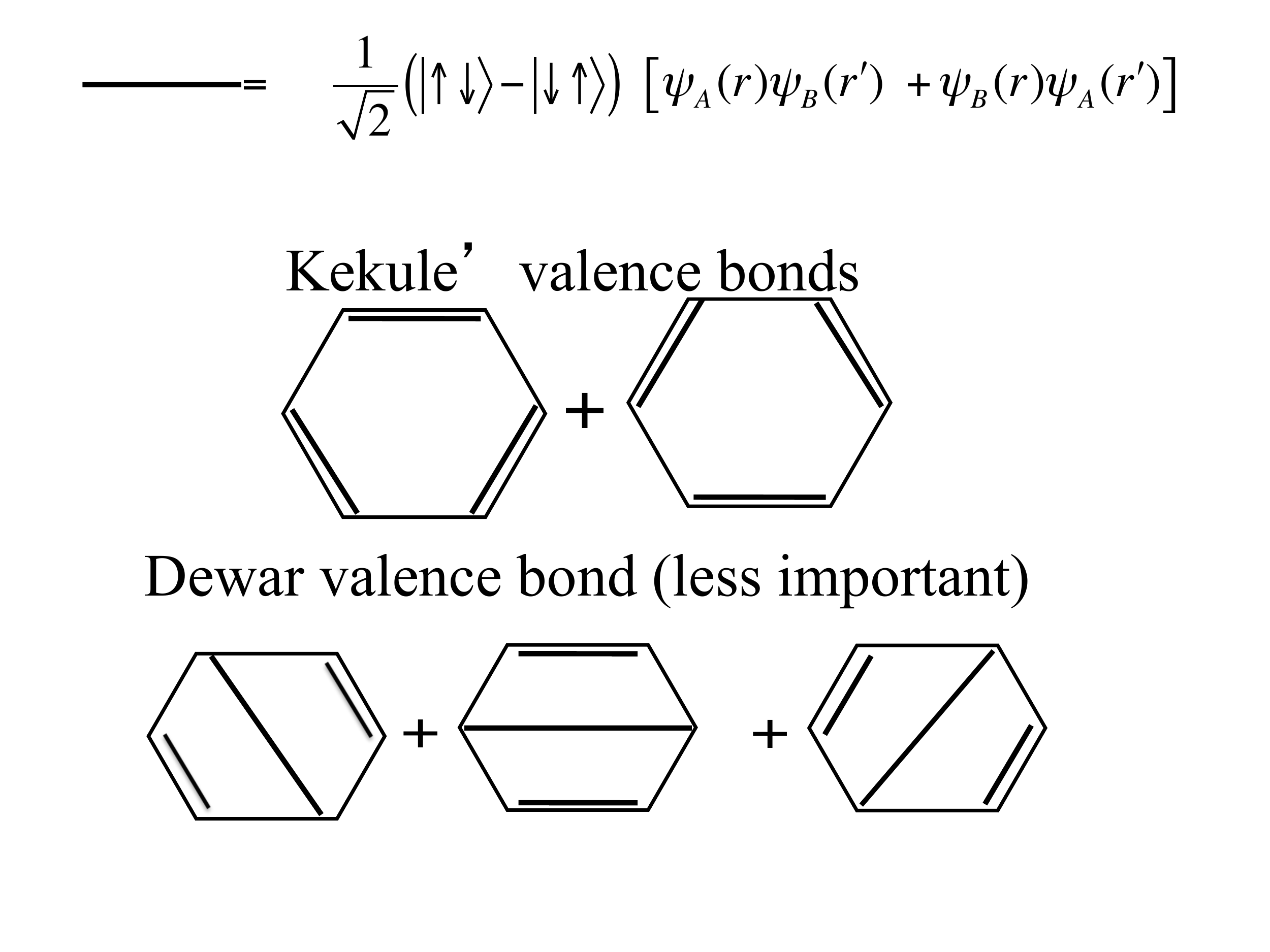}
\caption{Kekul\'e and Dewar contributions to the 
 resonance valence bond energy in the benzene molecule. Bold lines schematically 
indicate a two electron singlet joining the corresponding Carbon atoms located at the vertices of the hexagon. }
\label{benzene}       
\end{figure}

Suppose to extend this picture to a more complex molecule like benzene, a planar molecule 
made of six Hydrogen atoms and six Carbon atoms (chemical formula $C_6 H_6$), placed at the vertices of 
 an ideal hexagon. In this molecule 
there is only one $p_z$ orbital per Carbon atom 
 pointing in the $z$ direction perpendicular to the molecular plane. 
This single particle state  can be occupied only by one valence electron per Carbon  as, in first approximation,  all the 
remaining 36 electrons of the molecule are well described by a single Slater determinant.
A first success of the original RVB theory was to identify the most important 
two-electrons singlet bonds in this molecule, that are shown in Fig.(\ref{benzene}). 
Each structure is obtained by joining two-electrons singlet bonds among neighboring 
atoms in the hexagon, indicated here by bold lines.
Bold lines cannot superpose, because otherwise two valence electrons with opposite spins 
 have to occupy the same $p_z$ orbital on the same atom, implying an high energetic cost
due to  the strong Coulomb repulsion.
Therefore, only by the superposition of inequivalent structures the electronic wave function can be 
symmetric under rotation by $60$ degrees, and  
the corresponding energy gain is called the "resonance valence 
bond energy".
In Fig.(\ref{benzene}),
for each valence bond structure, represented by an hexagon, 
we can expand the corresponding three singlet bonds 
 in terms of orbital functions. 
We thus obtain a different Slater determinant for any 
 of the possible $2^3=8$  spin configurations obtained  in each hexagon.
 Then if we count the total number 
of determinants, corresponding to the five Kekule' and Dewar structures, we conclude that already in this simple molecule we need fourty determinants to represent this RVB wave function. 
It is easy to realize that in a complex system containing several atoms, the number 
of such determinants  grows exponentially with the number of atoms $N_A$,  at least because 
a single valence bond, where singlets between different atoms  are drawn in the same 
way as for a single hexagon in Fig.(\ref{benzene}),  
 requires  at least $2^{N_A/2}$ determinants, and the method cannot be effectively 
applied to realistic systems containing several atoms. 
Basically for this simple reason the  Pauling RVB approach was soon abandoned, and 
the single-determinant technique based on molecular orbitals became popular for its simplicity 
and effectiveness. 

\begin{figure}[b]
\includegraphics[scale=.2]{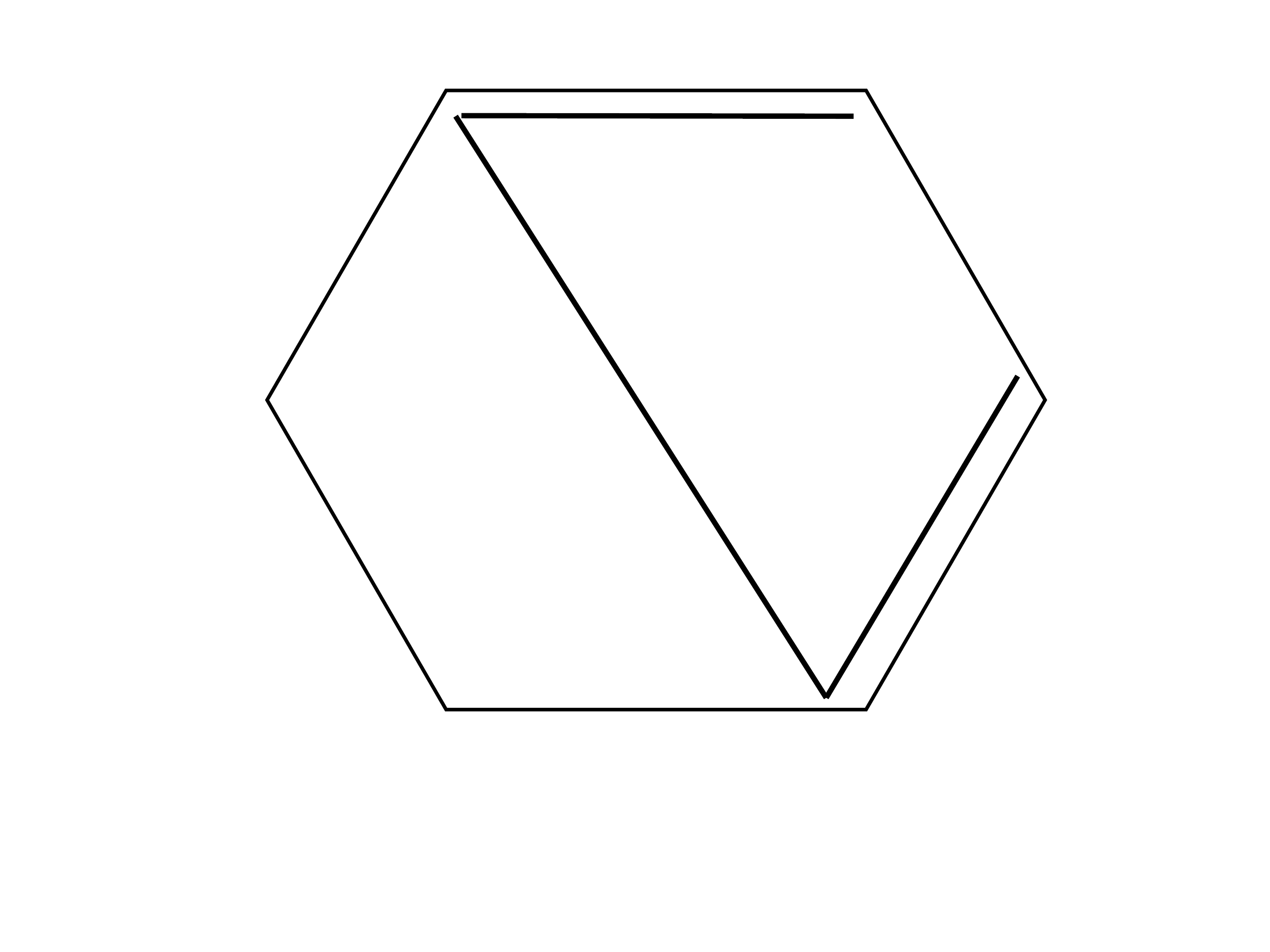}
\caption{An example of a valence bond configuration in benzene that is forbidden by the Jastrow 
factor $J(x) \sim 0$ because two pairs of electrons occupy the same valence orbitals in the two 
Carbon sites connected by the long bond.}
\label{supervb}       
\end{figure}

\begin{figure}[b]
\includegraphics[scale=.4]{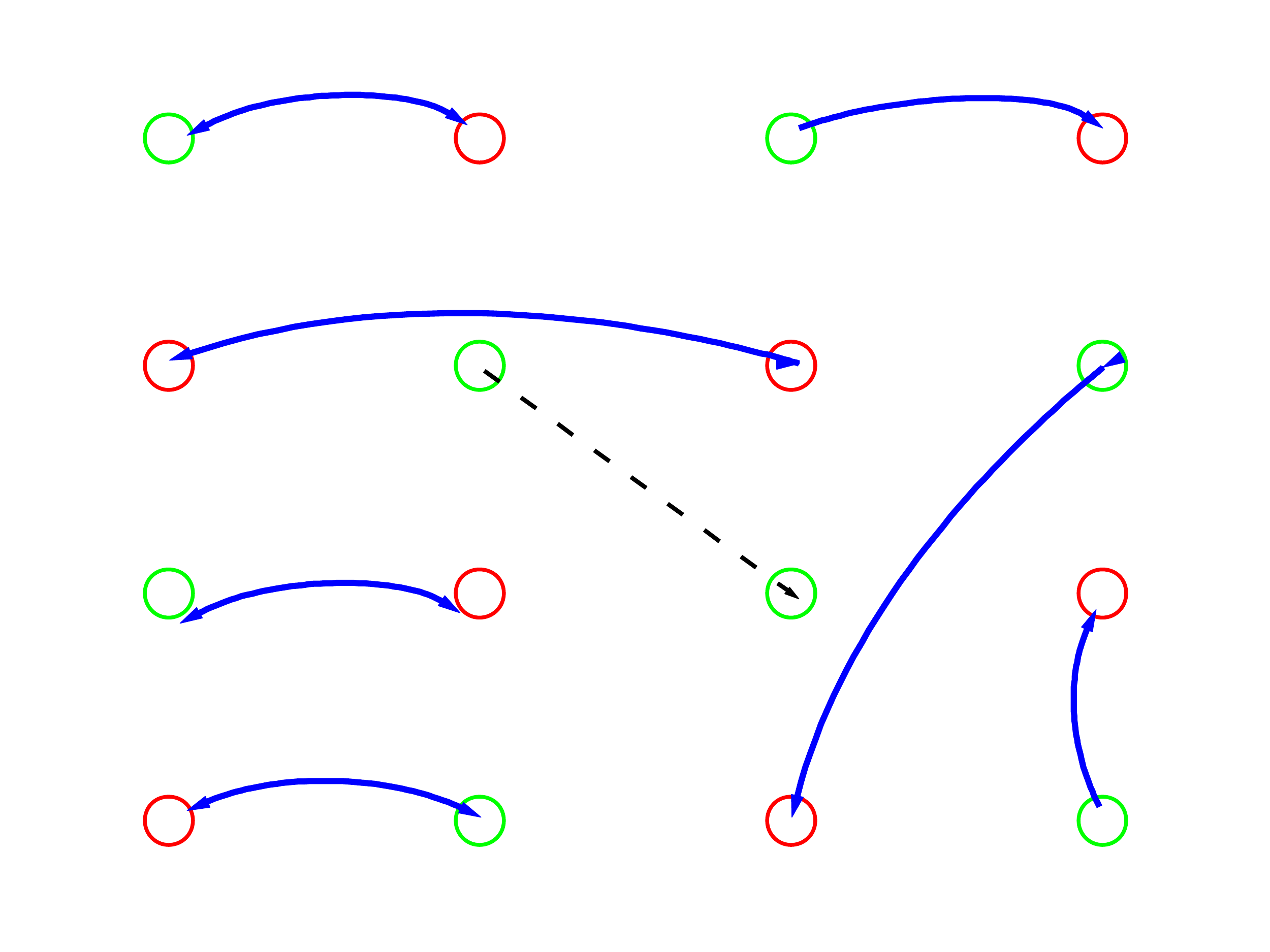}
\caption{ Schematic example of the mechanism of High-temperature superconductivity (HTc) within the Resonating Valence Bond theory. In the insulator each valence bond structure covers all valence orbital sites and charge cannot propagate in the insulator. Upon small doping  empty sites appear in the lattice 
(connected here by a dashed line). Such holes  
can freely  move in this soup of preformed electron pairs, leading to HTc supercurrent flow.
 For clarity the two antiferromagnetic sublattices are  indicated by green and red circles.
 }
\label{htc}       
\end{figure}

After several decades the RVB theory for High-temperature superconductors was introduced 
by Anderson in 1987 \cite{pwanderson}. 
The key ingredient in this theory is that the singlet pairing function $f$  describing 
a superconductor, when expressed in real space, is exactly equivalent to the function $f$ 
already introduced for a chemical bond.
Indeed the wave function of a superconductor for a system with fixed number $N$ of electrons can 
be described by the so-called antisymmetrized geminal product (AGP) wave function:
\begin{eqnarray} \label{agppairing}
F(x) &=&F(\vec r_1, \sigma_1; \vec r_2, \sigma_2; \cdots; \vec r_N, \sigma_N ) \nonumber  \\
 &=& {\cal A} f(\vec r_1,\sigma_1; \vec r_2,\sigma_2) 
              f(\vec r_3,\sigma_3; \vec r_4,\sigma_4) 
              \cdots 
              f(\vec r_{N-1},\sigma_{N-1}; \vec r_N,\sigma_N ) 
\end{eqnarray} 
where ${\cal A}$ is the antisymmetrization operator over all particle permutations, 
and $x$ indicate conventionally a real space configuration, where all $N$ electrons have 
 definite positions and spins $\sigma_i$ along the $z$ spin quantization axis.
If, for instance, we take a pairing function for the benzene molecule localized in 
a given Kekule' structure, by applying the above antisymmetryzation we obtain almost all
valence bonds, but unfortunately something more.
Indeed there is nothing that forbid the above expansion to generate a diagram like 
the one in Fig.(\ref{supervb}).
, namely the bonds can superpose, violating the simple and most important constraint of the 
RVB theory, that two electrons cannot occupy the same valence $p_z$ orbital due to the 
strong Coulomb repulsion.
Anderson's idea is basically that an explicit, correlated factor $J(x)>0$ can avoid these energetically 
expensive configurations, and achieves the target of an RVB wave function, built by a superconducting, namely an AGP function:
\begin{equation} \label{newrvb}
 \Psi_{NewRVB} = J(x) F(x)
\end{equation}
What have we gained with this new definition?
\begin{framed}
The most important achievement was to understand a possible mechanism of superconductivity.
High-T$_c$ superconductors are close to Mott insulators well-described by an RVB wave function
where the pairing function $f$ has $d-$wave symmetry and the phase coherence implied by 
the $F(x)$ alone is instead suppressed by the correlation factor $J(x)$. 
As it is shown in Fig.(\ref{htc}), by a small amount of doping 
these preformed pairs allow charge propagation  and lead to a faithful description of an 
High-temperature superconductor with a finite $d-$wave off-diagonal 
long range order.
\end{framed}

In the following instead we use the wave function paradigm given in Eq.(\ref{newrvb}) 
just as a convenient numerical ansatz to represent a RVB wave function.
Indeed, as it will be shown in the following, in a given electronic configuration $x$, both 
the  AGP $F(x)$ and the Jastrow factor $J(x)$ can be computed with a reasonable number of operations,
namely scaling at most as the third power of the number of electrons.
\begin{framed}
It is important here to emphasize that, once the Jastrow factor is taken to satisfy the 
constraint of no doubly-occupied valence states, we need only one pairing function to describe  all 
the Kekule' and Dewar structures because $f$ can be taken as the sum over all the six nearest-neighbor (for Kekule') Carbon-Carbon bonds  plus a small weight of the six largest 
distance ones (for representing the Dewar structures).
As anticipated this means that correlation is localized in space, and its overall effect in 
a complex structure is obtained by simply summing up in $f$ all these space independent contributions. 
After that we need only the computation of a single determinant $F(x)$ and a simple 
Jastrow factor $J(x)$, for  evaluating a wave function described in principle by an exponentially 
large number of Slater determinants.
\end{framed}

\section{The Wave Function}
\label{sec:2}

The wave function $\Psi_{NewRVB}$, that we have considered in this work, is written as the product of an antisymmetric (fermionic) function $F(x)$, and a symmetric (bosonic) exponential function $J(x)=e^u$:
where both $F$ and the Jastrow factor $J$ depend on the 
spatial and spin coordinates $\textbf{x}_i=(\textbf{r}_i,\sigma_i)$ of the $N$ electrons in the system, 
$x=\{ \textbf{x}_i \}_{i=1,\ldots,N}$.
Once the pairing function $f$ and the correlation factor $u$ are defined, the value of the total 
wave function 
$\Psi_{NewRVB}(x)$ 
can be computed efficiently on a given configuration
containing $N/2$ electron pairs of opposite spin electrons $\vec r_i^\uparrow,\vec r_i^\downarrow$:
\begin{equation}
\Psi_{NewRVB}(x) =
  \exp ( \sum\limits_{i<j} u(\vec r_i,\vec r_j) )  {\rm Det}  f(\vec r_i^\uparrow,\vec r_j^\downarrow) 
\end{equation}
After that, a standard variational quantum Monte Carlo approach is possible in order to compute the 
expectation value of the energy and correlation functions, with a reasonable computational time, 
scaling very well with the number of electrons, i.e. $N^3$.  
This is described later in more detail. 
For a complementary view on constructing correlated wave functions for quantum chemistry, see the Chap. ``Tensor Product Approximation (DMRG) and Coupled Cluster Method in Quantum Chemistry'' by Legeza et al.
For the time being it is important to emphasize that,   
as described in the introduction, the Jastrow term is chosen as to 
employ the local projection of no doubly-occupied valence electrons, that should be 
a consequence of an accurate energy minimization. 
 On the other hand, for the same reason, the parametrization of the pairing function $f$, 
the basic ingredient of $F(x)$, has to be described in detail in real space, in order to represent 
each correlated singlet bond.  
The two pairing functions $u$ and $f$ can be conveniently expanded  by using two different set of 
atomic orbitals. To this purpose, we consider  
an atomic basis $\{ \phi_\mu(\vec r) \}$, where  each element $\phi_\mu$ is localized around 
an atomic center $\vec R_\mu$ (obviously several elements may refer to the same atomic center). Then the pairing function $f$ can generally be written as:
\begin{equation} \label{agpwf}
 f(\vec r,\vec r^\prime) = \sum\limits_{\mu,\nu} f_{\mu\nu} \phi_\mu(\vec r) \phi_\nu(\vec r^\prime) 
\end{equation}
where $f_{\mu\nu}$ is now a symmetric finite matrix, satisfying the following important properties:
\begin{framed}
\begin{enumerate}
\item if the atomic basis is large enough and reaches completeness, it is possible to represent also 
the two particle functions $f$ and $u$ in a complete way.
\item the atomic basis $\phi_\mu$ is not necessarily orthonormal.
Actually, for practical purposes, it is convenient to choose simple, e.g., Gaussian or Slater, localized 
orbitals, without any orthogonalization constraint. 
\end{enumerate}
\end{framed}

Analogously, also the correlation term $u$ can be expanded in a different set of atomic orbitals.
However, in order to speed up the convergence to the complete basis set (CBS) limit, or in other words to
parametrize satisfactorily this Jastrow term within a small basis, it is important to fullfill the so 
called cusp conditions, so that all the singular behavior of the function $u$ when $\vec r\to \vec r^\prime$ 
(electron-electron) or when 
$\vec r \to \vec R_a$ (electron-ion) 
are satisfied exactly, namely 
$u(\vec r,\vec r^\prime) \sim 1/2 |\vec r-\vec r^\prime| $  and $ 
u(\vec r,\vec r^\prime) \sim  -Z_a |\vec r - \vec R_a|$,
respectively. 
Here $R_a$ and $Z_a$  indicate the  atomic positions and the corresponding atomic number $Z_a$ of the electronic system considered, respectively.
The general form of the Jastrow correlation $u$ is therefore written in the following form:
\begin{equation} \label{jwf}
u(\vec r,\vec r^\prime) = 
  u_{ee} ( | \vec r - \vec r^\prime| ) 
  + (u_{ei} (\vec r)+u_{ei}(\vec r^\prime) ) 
  + \sum\limits_{\mu,\nu} u_{\mu\nu} \chi_\mu(\vec r) \chi_\nu(\vec r^\prime) 
\end{equation}
where $u_{ee}$ and $u_{ei}$ are simple functions  satisfing  the electron-electron 
and electron-ion cusp conditions and are reported elsewhere\cite{Marchi:2009p12614}.
In order to allow a general and complete description 
of the latter one-body term, it is also assumed that one 
orbital in the above expansion is just constant and identically one, say for $\mu,\nu=0$.
Then, it is simple to realize that this term just renormalize $u_{ei}$ by:
\begin{equation}
u_{ei}(\vec r) \to u_{ei}(\vec r) + 2 \sum_{\nu\ne 0} u_{0\nu} \chi_\nu(\vec r)
\end{equation}
so that, for a sufficiently large basis set,  both the single-body  and the two-body dependency of the 
Jastrow factor can be represented with  an arbitrary degree of accuracy and detail.

$\{ \chi_\mu({\bf r}) \}$ is also a localized basis set, exactly of the same form as $\phi_\mu$ used for expanding the pairing 
function $f$. In this case, however, it is convenient to use a different set of orbitals, and usually 
a much smaller basis dimension is necessary to obtain converged results, at least in the energy 
differences and for the relevant chemical properties.

In the following  we will provide a synthetic description of the atomic orbitals that are used to write both the determinantal and the Jastrow parts of the wave function.

\subsection{Atomic Orbitals}\label{sec.orbitals}

A  generic atomic orbital $\phi_{\mu}$ centered at the position $\textbf{R}_\mu$ 
is  written in terms of the radial vector 
$\textbf{r}- \textbf{R}_{\mu} $
connecting the position $\textbf{R}_\mu=\textbf{R}_a$ of nucleus $a$ to the position $\textbf{r}$ of an electron.
Hereafter the atomic index $a$ will be neglected in order to simplify the notation.
Of coarse there are several atomic orbitals used to describe each atom $a$.

In this work we  consider the most general atomic 
orbital centered around the atomic position $\vec R_\mu$ that can be expanded in terms of simple 
elementary atomic orbitals. These elementary orbitals are determined by a radial part given by a simple Gaussian or Slater form,  
whereas their angular part is characterized  
by an angular momentum   $l$ and its projection $m$ along a given axis: 
\begin{equation}\label{eq.STO}
\phi^{STO}_{l,m}({\bf r};\zeta) \propto \|\vec r - \vec R_\mu \|^l e^{-\zeta \| \vec r-\vec R_\mu\|} Y_{l,m}(\Omega)
\end{equation}
and the Gaussian type orbitals (GTO):
\begin{equation}
\phi_{l,m}^{GTO}({\bf r};\zeta) \propto \|\vec r-\vec R_\mu\|^l e^{-\zeta \| \vec r-\vec R_\mu\|^2} Y_{l,m}(\Omega)
\end{equation}
where $Y_{l,m}(\Omega)$ is a real spherical harmonic centered around $\vec R_\mu$.
The proportionality constant is fixed by the normalization and depends on the parameter $\zeta$.
Other parametric forms for the atomic orbitals exist, see for instance \cite{Petruzielo:2011p24345}, but have not been used in this work.

As discussed previously, the electron-ion  cusp condition is satisfied by the  term $u_{ei}$ included in the Jastrow term. 
For this reason we need smooth atomic orbitals with no cusps at the nuclear positions. 
This is automatically satisfied by all the GTO and STO orbitals  described here, with the exception of 
 the $s$-orbital STO (i.e., $l=m=0$), that is smoothed 
as follows:
\begin{equation}\label{eq.STO.s}
\phi_{0,0}^{STO}({\bf r};\zeta) \propto (1+\zeta \| \vec r-\vec R_\mu\|) e^{-\zeta \| \vec r-\vec R_\mu\|} Y_{0,0}(\Omega)
\end{equation}
Observe that each elementary orbital  described here depends parametrically only on the exponent 
$\zeta$.

The most general atomic orbital $\phi_\mu (\vec r)$ can be  expanded in terms of elementary orbitals
as follows:
\begin{equation}
\phi_{\mu}(\textbf{r}) = \sum\limits_{l,k,m} c^k_{l,m} \phi^{X_{k,l}}_{l,m}( \vec r,\zeta_{k,l}) ,
\end{equation}
namely it may contain elementary functions corresponding to different angular momenta, 
different types Slater ($X_{k,l}=STO$) or Gaussian ($X_{k,l}=GTO$), and different exponents 
$\zeta_{k,l}$. 
Usually in quantum chemistry methods, this type of operation is called contraction and is often 
adopted to reduce the atomic basis dimension for the description of strongly bound atomic 
orbitals (e.g., 1s). Therefore, it is not  common to hybridize different angular momenta.

In our approach instead, in order to describe the wave function with an affordable number 
of variational parameters, it is crucial to reduce the atomic 
basis dimension $D$ as much as possible, because the number of variational parameters 
(mostly given by the number of  matrix elements  $f_{\mu \nu}, u_{\mu \nu}$ in Eqs.\ref{agpwf},\ref{jwf}) 
is proportional to the basis dimension square, namely, $D (D+1)/2$, only for $f$. 
It is therefore extremely important to reduce this number $D$  by optimizing the independent atomic 
orbitals in a large primitive basis of elementary functions.  In this way a very small 
number of contracted orbitals -- referred to as ``hybrid'' orbitals here -- 
are necessary to reach converged variational results.

\subsection{Molecular Orbitals}

A generic orbital of a Slater Determinant, for instance,  within  Hartree-Fock theory, 
can be expanded in terms of atomic orbitals, namely, it has components spread  over all the atoms 
of the  system considered.
These orbitals are usually called molecular orbitals (MO). 

In the following sections we  consider other functional forms for the determinantal part of the wave function.
The relation between the AGP and  those other wave functions can be easily understood by rewriting the pairing function $f( \textbf{r}_{i},\textbf{r}_{j} )$ in the following equivalent way. 
First of all we diagonalize the matrix ${\bf f}$, whose  elements are the $f_{\mu\nu}$, by taking  
into account that the atomic orbitals are not necessarily orthogonal each other, namely the matrix $\bf S$ has elements $S_{\mu,\nu} = \langle \phi_\mu| \phi_\nu \rangle \ne \delta_{\mu\nu}$.
 This can be done by using a standard generalized diagonalization:
\begin{equation} \label{gendiag}
{\bf f S  P} = {\bf P} \Lambda.
\end{equation} 
where the generalized eigenvectors of $\bf f$ define each column of the 
 matrix $\bf P$, whereas  
$\Lambda$ is a diagonal matrix containing the corresponding generalized eigenvalues $\lambda_\alpha$. 
Here, for notational convenience,
the non-vanishing eigenvalues $\lambda_\alpha$  are sorted in ascending  order, 
according to their absolute values:
$ |\lambda_1| \le |\lambda_2| \le \ldots \le |\lambda_n |$.
Thus, from $ {\bf P}^T {\bf S P} = {\bf I}$, by right-multiplying both sides of Eq.(\ref{gendiag})  for the 
matrix ${\bf P}^T= ({\bf SP})^{-1}$ we obtain   ${\bf f} = {\bf P}  \Lambda   {\bf P}^T$.
Then, by substituting it in \ref{agpwf}, we finally obtain that 
the pairing function can be also written as 
\begin{equation}
f\left( \textbf{r},\textbf{r}^{\prime} \right) =
  \sum_{\alpha=1}^{n}  
    \lambda_{\alpha} 
    \Phi_{\alpha}\left(\textbf{r}\right)
    \Phi_{\alpha}\left(\textbf{r}^\prime \right) ,
\label{equ:phiG_2}
\end{equation} 
namely, in terms of generalized orthonormal MOs:
\begin{equation}
\Phi_\alpha(\textbf{r}) = \sum\limits_i c_{\alpha,i}  \phi_i(\textbf{r}). 
\end{equation}
Notice that, if the number of non-zero eigenvalues $\lambda_\alpha$ is exactly equal to the number 
of pairs $N/2$, the antisymmetrization in Eq.(\ref{agppairing}) projects out 
only a single Slater determinant, and the molecular orbitals coincide in this case with the standard 
 ones.
From this point of view it is transparent that this wave function can improve the Hartee-Fock 
single-determinant picture, especially when, as discussed in the introduction, it is  combined with the Jastrow factor.

\begin{framed}
We remark here, that, even when the AGP is exactly equivalent to a Slater determinant ($n=N/2$), 
the combined optimization of the Jastrow factor and the molecular orbitals may lead to a qualitatively 
different wave function, namely with different chemical and physical properties.
In such case it is fair to consider  this kind of wave function as an RVB and therefore it will be  also indicated in the following with the  nRVB acronym. 
\end{framed}

\subsection{New RVB Wave Function}\label{sec.AGPn}

It has been proved~\cite{Sorella:2007p12646,Neuscamman:2012hm} that a Jastrow correlated AGP function, the 
new RVB  (nRVB),   satisfies the size-consistency of singlet fragments, 
namely the energy of the system is equal to the sum of the energies of the fragments, when the 
fragments are at very large distance.
This property holds provided that the Jastrow term is flexible enough.

The fully optimized  wave function with $n=N/2$, that will be denoted hereafter by JHF,
corresponds to a Jastrow correlated   Hartree-Fock wave function, and it
provides an accurate description of atoms\cite{Foulkes:2001p19717,Marchi:2009p12614}, with 
more than $90\%$ of the correlation energy.
The idea here is to use the larger variational freedom given by the 
 nRVB ansatz for $n>N/2$, only 
 to improve  the chemical bond, 
without requiring an irrelevant improvement of the isolated atoms.  
A natural criterion for restricting the number $n$ of MOs to a minimal number in a molecule or 
any electronic system, is to require that, when the atoms are at large distances, we cannot obtain an energy below the sum of the JHF atomic energies.
The number $n^*$ of MOs defined in this way is  therefore determined by the obvious requirement that:
\begin{equation}
n^* \ge \sum_a^M {\rm Max} ( N^{\uparrow}_a, N^{\downarrow}_a ) 
\end{equation}
 as we need at least one atomic orbital for each pair with opposite spin electrons and  each unpaired 
electron, where $N^\uparrow_a$ ($N^\downarrow_a$) is the maximum number of spin-up (spin-down) electrons 
contained in the atom $a$.
A rank $n$ of the nRVB larger than $n^*$ or equal to  $n^*$ will be sufficient  to build uncoupled atomic wave functions at large distance by means of a factorized $f_{\mu\nu}$ in terms of block-diagonal atomic contribution.
For further details and for a discussion of polarized systems see 
Ref.~\cite{Marchi:2009p12614}.

\begin{framed}
 This constraint on the number $n$ of molecular orbitals is 
not only useful to reduce the number of variational parameters but it is extremely important to 
improve the description of the chemical bond.
In fact, we have reported  a number of cases~\cite{Marchi:2009p12614}  where, 
by increasing $n$ to a value much larger than $n^*$,  the accuracy in the 
description of equilibrium properties and chemical properties drastically deteriorates.
A larger value of $n$ guarantees a lower total energy because, by the variational principle, to a larger variational freedom corresponds a lower total energy. 
However a larger $n$ may improve too much the atomic description-depending on high energy details-at the price to deteriorate the low-energy chemical properties (see next section).
\end{framed}


\section{The Variational  Quantum Monte Carlo Method}
In order to deal with a correlated wave function, with an explicit Jastrow correlation term, 
the  simplest  and most efficient method is by far the variational quantum Monte Carlo (VMC), 
introduced long time ago by McMillan in 1965 for the first calculations on $He_4$.  
After that,  progress has been made. In particular, 
very useful schemes  for the optimization of several 
variational parameters in VMC were successfully developed in the last decades~\cite{Sorella:2005p14143,Umrigar:2007p12662}.
Other approaches based on the idea of QMC are treated in this book: in Chap. ?Levy-Lieb Principle Meets Quantum Monte Carlo? by Delle Site; in Chap. "Mathematical Perspective on Quantum Monte Carlo Methods" by Cances;
in Chap. "Electronic Structure Calculations with LDA+DMFT" by Pavarini. 
In particular, in the chapter of Cances, the mathematical formalization of VMC is discussed. 

In the early days, only few variational parameters were introduced to describe any electronic system, as for instance Helium. 
Nowadays it is common to work with several thousands of them, that allows 
for a faithful description
and  in principle to reach the complete basis set limit 
of a given ansatz~\cite{Sorella:2007p12646}.
The method is based on a stochastic evaluation of the total energy and its 
derivatives (e.g., with respect to variational parameters and/or ionic positions, namely 
the atomic forces~\cite{Attaccalite:2008p12639,Sorella:2010p23644}). An appropriate  Markov chain in configuration space generates 
configurations distributed according to the wave function square 
$\left| \Psi_{newRVB}(x) \right|^2$. 
 The expectation values of any physically interesting 
observable $\cal O$ is evaluated  by calculation of the sampling mean $\left< O \right>$  for the corresponding random variable $O$, 
defined on a given configuration.  The corresponding standard deviation $\sigma_{\left< O \right>}$, converges to zero as the inverse square root of the simulation length, and, nowadays, 
by means of supercomputers,  can be reduced to the desired level of accuracy even for systems containing several 
hundred atoms. 

At the end of this section, for clarity, 
it is worth to single out the random variable associated to the energy, the so-called 
local energy:
\begin{equation} \label{eloc}
e_L(x) ={  \langle x | H | \Psi_{newRVB} \rangle \over \langle x | \Psi_{newRVB} \rangle }
\end{equation}
All the other random variables can be  obtained by replacing in the above expression the hamiltonian $H$ with the corresponding physical operator $O$.

\label{sec:3}
\section{ Examples on Beryllium Dimer  and Water Molecule }

\begin{figure}[b]
\hspace*{-1.1truecm}
\begin{center}
\includegraphics[scale=.6]{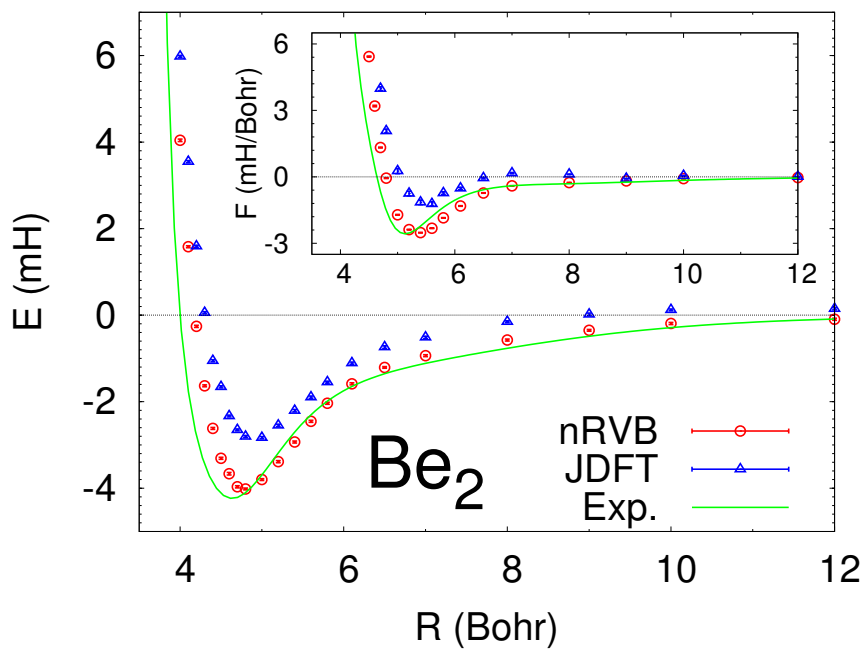}
\end{center}
\caption{Dispersion energy for $Be_2$ molecule. The nRVB calculation refers to the full optimization 
of a simple  Slater determinant in presence of the Jastrow term. In the JDFT calculation only 
the Jastrow factor is optimized. The experimental curve is derived by direct evaluation of vibrational 
frequencies\cite{Be2exp}. In the inset the atomic force is  displayed.}
\label{be2}       
\end{figure}

In this section we present one of the most successful examples, among many of them, of 
the simultaneous optimization of the Jastrow and the determinantal part of the wave function. The Beryllium dimer is one of the most difficult molecules to describe 
by ab-initio methods.
The simplest Hartree-Fock does not predict  a stable molecule and, despite the fact the correct 
binding energy is very small ($\simeq 0.1 eV$),  
the  hybridization of 2s and 2p atomic orbitals together with a 
weak van der Waals (vdW)  interaction 
yield a rather short bond length.

Several attempts have been made, using DFT methods with sophisticated RPA treatments of the long-range interaction~\cite{BeGalli}, but so far a reasonable description
 of the bonding has  been obtained only by using an expansion in 
several (billion) determinants~\cite{fullCI,Be2revisited}.    
Fortunately, recently there was a very accurate experimental paper~\cite{Be2exp}, 
that has proven to be very useful to benchmark the various calculations.

In Fig.(\ref{be2}) we report two different variational Monte Carlo calculations: 
the most accurate (nRVB) is obtained by applying a Jastrow factor $u$ and a pairing function $f$ with $n=n^*=4$ 
molecular orbitals  (the basis set is $4s4p2d$ and  $5s2p1d$ respectively for $u$ and $f$), fully optimized 
within an accuracy of $0.01mH$;
whereas the less accurate one (JDFT) is obtained by using 
DFT (with standard LDA approximation) 
with a large VTZ basis ($11s11p2d1s$) for determinining the molecular orbitals, and only $u$ is fully optimized in the same basis used for the nRVB. 
We remark here that the small basis used in the nRVB case is already enough to reach an accuracy  on the total energy  within $0.1mH$. This is because the full optimization of the wave function  in presence of a large Jastrow factor, allows a rapid convergence, and requires therefore a minimal basis for $f$.  Indeed, despite the small basis, the 
nRVB total energy is about 3mH below the JDFT one in all the interatomic distances under consideration.

As it is shown in this picture, the simultaneous optimization of the Jastrow and 
the determinantal part of the wave function is fundamental for obtaining a good agreement with the experimental results. 
The most important ingredient  to describe this bond is clearly given by our accurate 
 Jastrow factor, which is capable to correctly describe  the long-range vdW interaction. 
Without the optimization of the determinantal part, however, the agreement with the experiments is rather poor, and  qualitatively wrong at large distance. 
This suggests that, in this difficult case, a good account of the correlation 
is possible with a single determinant only when there is a strong coupling between the Jastrow and the determinantal part, as in the RVB picture described in the introduction.
We remark that, our  results, are much 
better than our previous work~\cite{Marchi:2009p12614}, 
just because in the present case we have achieved the complete basis set (CBS) limit, with a more accurate Jastrow factor.

We also show here some very recent results for the water molecule, obtained by exploiting the efficiency in the representation of the nRVB by means of the general hybrid contracted 
orbitals defined in the previous section.
Within this approach, it is easy to obtain converged results for the total energy and in all the interesting physical properties of the molecule, such as the atomization energy, 
the dipole and the quadrupole tensor, the equilibrium structure and the vibrational frequencies~\cite{Zen:2013is}.
We see in Tab.\ref{table:water} that the JDFT calculation is rather accurate and it describes all these properties in a quantitative way.
In the water molecule, the physics of the RVB is certainly not as crucial as for Be$_2$.
However, also here
the larger variational freedom
allows us to improve significantly the agreement of the estimated quantities 
with the experiments or the estimated exact results.
The improvement is observed in particular  in the charge distribution (see the dipole in Tab.\ref{table:water}) and in the potential energy surface (see equilibrium and frequencies). 
It is important to emphasize that this improvement in the description of the water molecule is given for free in QMC, because the cost to optimize the simple 
Slater determinant in presence of the Jastrow factor is essentially the same as for the full nRVB wave function with no constraint on molecular orbitals.  

\label{sec:5}

\begin{table}
\caption{
Properties of water molecule obtained from VMC calculation with  JDFT, JHF and nRVB ansatz (with sufficiently large basis sets to be considered at convergence), in comparison with estimated exact results (from experiments or from very accurate calculations, see references).
The energy conserving pseudo potential of \cite{Burkatzki:2007p25447} is used to describe the the two core electrons of the oxygen atom.
The values in parenthesis represent the estimated error of the value (that is a stochastic error for the QMC calculations).
The quantities considered are the dipole (D), the quadrupole ($Q_{xx}$, $Q_{yy}$, $Q_{zz}$, where the orientation of the molecule is that described in \cite{Zen:2013is} and relative to the center of mass reference), the equilibrium configuration ($r_{OH}$,$\alpha_{HOH}$), the fundamental vibrational frequencies ($\nu_1$, $\nu_2$, $\nu_3$) and the atomization energy AE. 
The atomization energy for the water molecule, calculated as $E_{H_2O} - (E_{O}+2 E_{H})$,
where $E_{H_2O}$ is the energy calculated for the water molecule, 
$E_{O}$ for  the oxygen atom, and the hydrogen one $E_H$ is equal to 0.5H.
}  
\label{table:water}
\begin{tabular}{ l c c c c  c c   }  
\hline\hline
        & D$[Deb]$   & $Q_{xx}[Deb\AA]$     & $r_{OH}$$[\AA]$       & $\nu_1 [cm^{-1}]$   &  $E_{H_2O}$$[H]$  & AE$[H]$    \\ 
        &            & $Q_{yy}[Deb\AA]$     & $\alpha_{HOH}$$[deg]$ & $\nu_2 [cm^{-1}]$   &  $E_O$$[H]$   \\ 
        &            & $Q_{zz}[Deb\AA]$     &                       & $\nu_3 [cm^{-1}]$   \\ 
\hline \hline
 JDFT   & 1.9059(8)   & 2.5796(9)   & 0.95497(3)   & 3693(2)    & -17.2455(2)      & 0.3710(3)      \\ 
        &             & -0.1551(9)  & 104.49(2)    & 1610(1)    & -15.8744(2) \\
        &             & -2.4245(9)  &              & 3787(3) \\
\hline
 JHF    & 1.8907(7)   & 2.5676(8)   & 0.95426(3)   & 3702(3)   & -17.2471(1) & 0.3702(2)     \\ 
        &             & -0.1419(8)  & 104.74(1)    & 1617(1)   & -15.8769(1)   \\
        &             & -2.4256(8)  &              & 3794(2) \\
\hline
 nRVB   & 1.8648(6)   & 2.5740(7)   & 0.95550(4)   & 3677(2)   & -17.25383(4)   & 0.3700(2)      \\ 
        &             & -0.1500(7)  & 104.41(1)    & 1613.3(6) & -15.8838(2)    \\
        &             & -2.4240(7)  &              & 3772(2) \\
\hline
 exact$^a$  & 1.8546(6)   & 2.63(2) & 0.95721(30)   & 3656.65    & -76.438   & 0.3707        \\ 
        &             & -0.13(3)    & 104.522(50)   & 1594.59    & -75.0673 \\
        &             & -2.50(2)    &              &  3755.79 \\

\hline \hline

\multicolumn{7}{ p{9cm} }{ 
$^a$ 
D from \cite{Clough:1973bh}, 
Q from \cite{Verhoeven:1970jd},
$r_{OH}$, $\alpha_{HOH}$ and $\nu$ from \cite{Benedict:1956id}, 
$E_{H_2O}$ from \cite{Feller:1987dm},
$E_O$ from \cite{Chakravorty:1993gg}
and $AE$ as difference.
} \\
\\
\end{tabular}
\end{table}

\section{Conclusions}
In this chapter we have described in a simple, yet complete, way the main reasons why the 
new RVB variational ansatz opens a new frontier for  electronic simulations.
Until now we have lived in the DFT era, where the detailed and often subtle (e.g., 
the long range interactions such as the van Der Waals one) correlations  are assumed to 
be implicitly described by a function of one electron coordinate, the local density $n(\vec r)$ 
(see in particular chapters of Giringhelli, von Lilienfeld and Watermann et al.).
In this approach instead the explicit many-body correlation is determined, often quite accurately, 
by two functions $f$ and $u$ of two electronic coordinates, 
the pairing function and the Jastrow correlation term, respectively. 

\begin{framed}
It is important to emphasize at this point the crucial role played by the mutual interplay 
of the Jastrow factor $u$ and the pairing function $f$. 
In a clean  uncorrelated system ($u=0$), the pairing function represents a metal if it is enough 
long range $|f (r,r^\prime)| \sim {1 \over |r-r^\prime|^2}$ since it behaves as the density matrix 
of a free electron gas for $|r-r^\prime| \gg 1$. 
When it decays exponentially or in general faster than $1/|r-r^\prime|^2$ it may represent  either 
a band insulator or a superconductor.  
In the latter case, the correlation factor can play a crucial role because it can  
suppress phase coherence  and give raise to a RVB insulator, that may be considered 
a ``correlation frustrated'' superconductor. 
This simple example shows that the nRVB paradigm, described in this chapter, is not only  useful 
to improve on a ``naive correlation'', but allows us to change the qualitative properties of an 
uncorrelated picture, opening the path to the description of new phases of matter.
\end{framed}

\section*{Acknowledgement}
We acknowledge Mariapia Marchi for sending us unpublished data about the beryllium dimer.


\end{document}